\newcommand{\cl}{\centerline}
\documentstyle[12pt]{article}
\begin{document}

\def\d{{\rm d}}
\begin{titlepage}
\hfill{CCUTH-94-05}\par
\hfill{IP-ASTP-13-94}\par
\setlength{\textwidth}{5.0in}
\setlength{\textheight}{7.5in}
\setlength{\parskip}{0.0in}
\setlength{\baselineskip}{18.2pt}
\vfill
\cl{\large{{\bf Extraction of $V_{ub}$ from the Decay $B\to \pi l \nu$}}}\par
\vskip 1.0cm
\cl{Hsiang-nan Li$^1$\ and Hoi-Lai Yu$^2$, }
\vskip 0.3cm
\cl{$^1$Department of Physics, National Chung-Cheng University,}
\cl{Chia-Yi, Taiwan, R.O.C.}
\vskip 0.3cm
\cl{$^2$Institute of Physics, Academia Sinica, Taipei, Taiwan, R.O.C.}\par
\vskip 0.3cm
\cl{\today }
\vskip 0.3cm
PACS numbers: 13.20He, 12.15Hh, 12.38Bx\par
\vskip 2.0 cm

\cl{\bf Abstract}

We develop the perturbative QCD formalism including Sudakov effects
for semi-leptonic $B$ meson decays. We evaluate the differential
decay rate of
$B\to \pi l \nu$, and find that the perturbative calculation is
reliable for the energy fraction of the pion above 0.3.
Combining predictions from the
soft pion theorems, we extract the value of the matrix element
$|V_{ub}|$ which is roughly $2.7\times 10^{-3}$.

\vfill
\end{titlepage}

\newpage

Exclusive semi-leptonic meson decays have been intensively studied
by several approaches, which provide the information of
the mixing angles in the Cabibbo-Kobayashi-Maskawa matrix
of the standard model.
Chiral symmetry and heavy quark symmetry have been applied to
the decays $K\to\pi l \nu$ \cite{LR} and
$B\to D l \nu$ \cite{N1}, respectively, from which model-independent
extraction of the matrix elements $|V_{us}|$ and
$|V_{cb}|$ is obtained.
For the heavy-to-light transition $B\to\pi l \nu$,
which gives reliable estimation of $|V_{ub}|$ \cite{IW},
neither of the above theories is appropriate.
Recently, this decay has been investigated
using the heavy quark effective theory (HQET) \cite{BLN},
in which only the normalization of the relevant form factors is
determined in terms of the soft pion relations.
An explicit evaluation of the decay $B\to\pi l \nu$
based on the perturbative QCD (PQCD) formalism has been proposed \cite{YS}.
However, it leads to results which are too small compared to current
experimental data.

In this letter we shall develop a modified PQCD approach to heavy meson
decays, which includes Sudakov effects \cite{BS,LS}.
These effects, arising from the all-order summation of large
radiative corrections, suppress contributions from the nonperturbative
region, and have been found to extend the
applicability of PQCD down to the energy scale of few GeV
in the study of elastic hadron form factors \cite{LS}.
We shall show that the PQCD approach is proper for the decay
$B\to\pi l \nu$, at least when the pion is energetic.

The amplitude of the considered process is written as
\begin{equation}
A(P_1,P_2)=\frac{G_F}{\sqrt{2}}V_{ub}{\bar \nu}\gamma_{\mu}(1-\gamma_5)l
\langle\pi (P_2)|{\bar u}\gamma^{\mu}b|B(P_1)\rangle\;,
\end{equation}
where $G_F$ is the Fermi coupling constant, and
$P_1$ and $P_2$ are the $B$ meson and pion momenta, respectively.
We start with the lowest-order factorization for the matrix
element $M^{\mu}=
\langle\pi (P_2)|{\bar u}\gamma^{\mu}b|B(P_1)\rangle$,
in which
the $b$ quark carries the momentum $P_1-k_1$,
and its light partner carries $k_1$. These momenta satisfy the
on-shell conditions $(P_1-k_1)^2=m_b^2$, $P_1^2=m_B^2$ and $k_1^2=0$,
$m_b$ and $m_B$ being the $b$ quark and $B$ meson masses,
respectively. We choose the Breit frame such that
$P_1^+=P_1^-=m_B/\sqrt{2}$, $P_2^+=\eta m_B/\sqrt{2}$
and all other components of $P_i$'s vanish, where $\eta$ is related to
the energy fraction of the pion by $P_2^0=\eta m_B/2$,
$0 \le \eta \le 1$.
$k_1$ has a minus component, defining the momentum fraction
$x_1=k_1^-/P_1^-$ in the $B$ meson wave function, and small
amount of transverse components ${\bf k}_{1T}$. The light valence quark
of the $B$ meson, after
absorbing the hard gluon, goes into the pion with the momentum
fraction $x_2$ and transverse momenta ${\bf k}_{2T}$.

We then consider how to group radiative corrections into
the basic factorization by locating
their leading momentum regions,
from which important contributions to loop integrals arise.
The important corrections are characterized by large single
logarithms, which are either collinear
or soft.
These two regions may overlap and give double
logarithms. It is known that single logarithms
can be summed to all orders using renormalization group (RG)
methods, while double logarithms must be organized by
the resummation technique \cite{CS}, which has been developed
in axial gauge $n\cdot A=0$, $n$ being the gauge
vector and $A$ the gauge field.

A careful analysis shows that reducible corrections on
the pion side produce double
logarithms with soft ones cancelled in the asymptotic region
$b\to 0$ \cite{LS}, $b$ being the conjugate variable to ${\bf k}_T$.
Hence, they are dominated by
collinear enhancements, and can be absorbed into the pion wave funtion.
Reducible corrections on the $B$ meson side also give double logarithms,
but the soft ones do not cancel
and the collinear ones are suppressed by the $B$ meson
wave function.
Therefore, these corrections can be absorbed into the $B$
meson wave function, which is also dominated by soft dynamics.
Irreducible corrections, with an extra gluon
connecting the pion and the $B$ meson,
give only soft divergences, which cancel
asymptotically. They are then absorbed
into a hard scattering amplitude.
Hence, the factorization picture holds after radiative
corrections are included.

With the above reasoning, the factorization formula
for $M^\mu$ in the transverse
configuration space can be written as,
\begin{eqnarray}
M^\mu&=&\int_0^1 \d x_{1}\d x_{2}\int
\frac{\d^2 {\bf b}_1}{(2\pi)^{2}}\frac{\d^2 {\bf b}_2}{(2\pi)^{2}}
\,{\cal P}_\pi(x_{2},{\bf b}_2,P_{2},\mu)
\nonumber \\
& &\times \,{\tilde H}^\mu(x_1,x_2,{\bf b}_1,{\bf b}_2,m,\mu)
\,{\cal P}_B(x_{1},{\bf b}_1,P_{1},\mu)\; ,
\label{fbd}
\end{eqnarray}
with ${\cal P}_\pi$ and ${\cal P}_B$
the pion and $B$ meson wave functions, respectively, and
${\tilde H}^{\mu}$
the Fourier transform of the hard scattering amplitude to $b$ space.
$\mu$ is the factorization and renormalization scale.
Both ${\cal P}_\pi$ and ${\cal P}_B$ contain double logarithms,
which will be summed up below.
The approximation $m_b\approx m_B =m=5.28$ GeV
has been made to simplify the analysis.

We outline the resummation procedure employed for ${\cal P}_\pi$
\cite{BS,LS}. If the double logarithms are grouped into an exponential
${\cal P}\sim \exp[-\ln m\ln (\ln m/\ln b)]$,
the problem will become simpler by
considering the derivative
$\d{\cal P}/\d\ln m=C{\cal P}$, where the coefficient $C$
contains only single logarithms, and can be treated by RG methods.
Because of the scale invariance of $n$ in the gluon propagator,
\begin{equation}
N^{\mu\nu}(q)=\frac{-i}{q^2}\left(g^{\mu\nu}-\frac{n^{\mu}q^{\nu}+
q^{\mu}n^{\nu}}{n\cdot q}+n^2\frac{q^{\mu}q^{\nu}}{(n\cdot q)^2}\right)\;,
\end{equation}
${\cal P}_\pi$ depends only on the ratio
$\nu_2^2=(P_2\cdot n)^2/n^2$.
It is then possible to relate the derivative
$\d{\cal P}_\pi/\d\ln P_2^+$ to $\d{\cal P}_\pi/\d n$,
which can be easily computed using the relations
$\d N^{\mu\nu}/\d n_{\alpha}=-
(N^{\mu\alpha}q^{\nu}+N^{\nu\alpha}q^{\mu})/q\cdot n$.
The momentum $q$ appearing at both ends of the differentiated gluon line
hints the application of the Ward identity. After adding together all
diagrams with different differentiated gluon lines,
we obtain an equation graphically described by fig.~1a,
in which the square vertex represents
$gT^a n^2P_2^\alpha/(P_2\cdot nq\cdot n)$, $T^a$ being related to the
Gell-Mann matrices $\lambda^a$ by $T^a=\lambda^a/2$.

Due to the factor $1/q\cdot n$ in the new vertex and
the nonvanishing of
$n^2$, the leading regions of $q$ are soft and ultraviolet,
in which fig.~1a can be factorized according to fig.~1b
to lowest order of $\alpha_s$.
The part on the left-hand side of the dashed line is exactly ${\cal P}_\pi$,
and that on the right-hand side is assigned to the coefficient $C$.
We introduce a function ${\cal K}$ to organize the soft
enhancements in the first two diagrams of fig.~1b, and ${\cal G}$
for the ultraviolet divergences of the other two diagrams. The
soft subtraction in ${\cal G}$ is to avoid double counting.
We then derive the differential equation,
\begin{eqnarray}
\frac{\d}{\d \ln P_2^+}{\cal P}_\pi=\left[\,2{\cal K}
(b_2\mu)+\frac{1}{2}\;{\cal G}(x_2\nu_2/\mu)+\frac{1}{2}\;{\cal G}(
(1-x_2)\nu_2/\mu)\right]{\cal P}_\pi\; ,
\label{qp}
\end{eqnarray}
where the functions ${\cal K}$ and ${\cal G}$ have been calculated
using RG methods \cite{BS}.
Solving eq.~(\ref{qp}), we obtain the solution
\begin{equation}
{\cal P}_\pi(x_2,b_2,P_2,\mu)=
\exp\left[-\sum_{\xi=x_2,\;1-x_2}\;s(\xi,b_2,\eta m)\right]
{\bar {\cal P}}_\pi(x_2,b_2,\mu)\; .
\label{sp}
\end{equation}
The explicit expression of the exponent $s(\xi,b,Q)$
has been obtained in \cite{LS}, and will not be shown here due to its
complexity.

The function ${\bar{\cal P}}_\pi$ still contains single
logarithms from ultraviolet divergences, which need to be
summed using RG methods \cite{BS}.
The large-$b$ behavior of ${\cal P}_\pi$ is then written as
\begin{eqnarray}
{\cal P}_\pi=\exp
\left[-\sum_{\xi=x_2,\;1-x_2}\;s(\xi,b_2,\eta m)
-2\int_{1/b_2}^{\mu} \frac{d\bar{\mu}}{\bar{\mu}}\gamma
_q(g(\bar{\mu}))\right]
\phi_\pi(x_2,b_2)\; ,
\label{eop}
\end{eqnarray}
$\gamma_q=-\alpha_s/\pi$ being the quark anomalous dimension
in axial gauge.
The $b_2$ dependence in $\phi_\pi$, which corresponds to the
intrinsic transverse momentum dependence of
the pion wave function \cite{JK}, will be neglected.

As to ${\cal P}_B$, the resummation of the double logarithms
is subtler. The self-energy correction to the massive $b$ quark,
giving only soft single logarithms, should be excluded.
On the other hand, ${\cal P}_B$ involves the invariants such as $P_1^2$,
which can not be related to $n$,
so that the technique of replacing $\d/\d m$ by $\d/\d n$ fails.
However, the above difficulities can be removed by applying
the eikonal approximation to the heavy quark line.
In the collinear region with the loop momentum $q$ parallel to $k_1$
and in the soft region, the $b$ quark line can be
replaced by an eikonal line:
\begin{equation}
\frac{(\not P_1-\not k_1+\not q+m)\gamma^{\alpha}}{(P_1-k_1+q)^2-m^2}
\approx \frac{(P_1-k_1)^{\alpha}}{(P_1-k_1)\cdot q}+{\rm R}\;,
\label{eik}
\end{equation}
where the remaining part ${\rm R}$
is less important.
The involved physics
is that a soft gluon or a gluon moving parallelly to $k_1$ can not explore
the details of the $b$ quark, and its dynamics can be factorized.

The first difficulty is then resolved, because self-energy
diagrams of an eikonal line are excluded by definition \cite{C}.
With the scale invariance of $P_1-k_1$ as shown
in eq.~(\ref{eik}), which is equivalent to the flavor symmetry in HQET,
$P_1-k_1$ does not lead to
a large scale, and the remaining large scale is only $k_1^-$.
Furthermore, an explicit lowest-order investigation shows that
${\cal P}_B$ depends only
on the single ratio $\nu_1^2=(k_1\cdot n)^2/n^2$,
and thus $\d/\d k_1^-$ can be replaced by $\d/\d n$ now.

Following the similar procedures to those for the pion, we obtain the
differential equation,
\begin{equation}
\frac{\d}{\d \ln k_1^-}{\cal P}_B=
\left[{\cal K}(b_1\mu)+\frac{1}{2}{\cal G}(\nu_1/\mu)\right]{\cal P}_B\;.
\label{dfb}
\end{equation}
It can be shown that
the functions ${\cal K}$ and ${\cal G}$
for the $B$ meson are exactly the same as those in eq.~(\ref{qp}).
It is then straightforward to derive the solution
\begin{eqnarray}
{\cal P}_B=\exp\left[-s(x_1,b_1,m)
-2\int_{1/b_1}^{\mu} \frac{d\bar{\mu}}{\bar{\mu}}\gamma
_q(g(\bar{\mu}))\right]
\phi_B(x_1,b_1)\;.
\label{sop}
\end{eqnarray}
At last, the RG analysis of ${\tilde H}^\mu$ gives
\begin{eqnarray}
{\tilde H}^\mu(x_i,b_i,m,\mu)
=\exp\left[-4\,\int_{\mu}^{t}\frac{\d\bar{\mu}}{\bar{\mu}}
\,\gamma_q(g(\bar{\mu}))\right]{\tilde H}^\mu(x_i,b_i,m,t)\;,
\label{13}
\end{eqnarray}
where $t$ is taken as the largest mass scale associated with the hard gluon,
$t=\max(\sqrt{x_1x_2\eta}m,1/b_1,1/b_2)$.
Having factorized all the large logarithms into the exponents,
we can then compute ${\tilde H}^\mu$ to $O(\alpha_s)$.

Substituting
eqs.~(\ref{eop}), (\ref{sop}) and (\ref{13}) into (\ref{fbd}),
we obtain the factorization formula for
$M^\mu=f_1P_1^\mu+f_2P_2^\mu$, where the form factors $f_1$ and
$f_2$ are given by
\begin{eqnarray}
f_1&=& -16\pi{\cal C}_Fm^2\int_{0}^{1}\d x_{1}\d x_{2}\,
\int_{0}^{\infty} b_1\d b_1 b_2\d b_2\,
\phi_B(x_1,b_1)\phi_\pi(x_{2})
\nonumber \\
& &\times x_1\eta h(x_1,x_2,b_1,b_2,m)
\exp[-S(x_i,b_i,m)]\;,
\label{f1}
\end{eqnarray}
and
\begin{eqnarray}
f_2&=& 16\pi{\cal C}_Fm^2\int_{0}^{1}\d x_{1}\d x_{2}\,
\int_{0}^{\infty} b_1\d b_1 b_2\d b_2\,
\phi_B(x_1,b_1)\phi_\pi(x_{2})
\nonumber \\
& &\times [x_1h(x_1,x_2,b_1,b_2,m)+(3-x_2\eta)
h(x_2,x_1,b_2,b_1,m)]
\nonumber \\
& &\times \exp[-S(x_i,b_i,m)]\;,
\label{f2}
\end{eqnarray}
respectively, with
\begin{eqnarray}
h(x_1,x_2,b_1,b_2,m)&=&
\alpha_{s}(t)K_{0}(\sqrt{x_{1}x_{2}\eta}mb_2)
\nonumber \\
& &\times [\theta(b_1-b_2)K_0(\sqrt{x_1\eta}mb_1)I_0(\sqrt{x_1\eta}mb_2)
\nonumber \\
& &\;\;\;\;
+\theta(b_2-b_1)K_0(\sqrt{x_1\eta}mb_2)I_0(\sqrt{x_1\eta}mb_1)]\;.
\label{dh}
\end{eqnarray}
${\cal C}_F$ is the color factor defined by ${\rm tr}(T^aT^a)=N_c
{\cal C}_F$, $N_c$ being the number of colors, and
$K_0$ and $I_0$ are the modified Bessel functions of order zero.
The complete Sudakov exponent $S$ is given by
\begin{eqnarray}
S(x_i,b_i,m)&=&s(x_1,b_1,m)+
s(x_2,b_2,\eta m)+s(1-x_2,b_2,\eta m)
\nonumber \\
& &-\frac{1}{\beta}\left[
\ln\frac{\ln(t/\Lambda)}{\ln(1/b_1\Lambda)}
+\ln\frac{\ln(t/\Lambda)}{\ln(1/b_2\Lambda)}\right]
\label{sda}
\end{eqnarray}
with $\beta=(33-2n_f)/12$, $n_f=4$ being the number of quark flavors, and
$\Lambda\equiv \Lambda_{\rm QCD}=100$ MeV here.
The Sudakov
factor $\exp(-S)$ decreases quickly in the large $b_i$ region and
vanishes as $b_i>1/\Lambda$.
We have kept the intrinsic transverse momentum dependence in $\phi_B$.
Since the $B$ meson wave function is dominated by soft dynamics,
this dependence is more important than that in the pion.

$\phi_\pi$ is chosen as
the Chernyak-Zhitnitsky model \cite{CZ},
\begin{equation}
\phi_\pi(x)=5\sqrt{3}f_\pi x(1-x)(1-2x)^2
\end{equation}
with $f_\pi=93$ MeV the pion decay constant.
For the $B$ meson wave function we consider \cite{S}
\begin{equation}
\Phi_B(x,{\bf k}_T)=N\left[C+\frac{m^2}{1-x}+\frac{{\bf k}_T^2}
{x(1-x)}\right]^{-2}\;,
\end{equation}
with $N=1.232$ GeV$^3$ and $C=-0.99993m^2$,
which are obtained from the normalizations
$\int\d x\int\d^2 {\bf k}_T\Phi_B=f_B/2\sqrt{3}$ and
$\int\d x\int\d^2 {\bf k}_T\Phi_B^2=1/2$,
$f_B=160$ MeV being the $B$ meson decay constant \cite{BLS}.
The Fourier transform of $\Phi_B$ gives
\begin{eqnarray}
\phi_B(x,b)=
\frac{\pi Nbx^2(1-x)^2}{\sqrt{m^2x+Cx(1-x)}}
K_1\left(\sqrt{m^2x+Cx(1-x)}b\right)\;,
\end{eqnarray}
$K_1$ being the modified Bessel function of order one.

Results of $f_1+f_2$
with $b_1$ and $b_2$ integrated up to the same cutoff $b_c$ are shown
in fig.~2a. We observe that at $\eta=0.3$
approximately 50\% of the contribution
to $f_1+f_2$ comes from the region with $\alpha_s(1/b_c)< 1$,
or equivalently, $b_c< 0.5/\Lambda$.
At $\eta=0.4$, 55\% of the contribution is accumulated in this
perturbative region. As $\eta=1$, perturbative contribution
has reached 75\%.
It implies that the modified PQCD analysis of the decay $B\to\pi l \nu$
in the range of $\eta > 0.3$
is relatively reliable according to the criteria given in \cite{LS}.
The differential decay rate for the
specific case $B^0\to\pi^- l^+\nu$ with vanishing
lepton masses is given by
\begin{eqnarray}
\frac{\d \Gamma}{\d \eta}\equiv |V_{ub}|^2R(\eta)
=|V_{ub}|^2\frac{G_F^2m^5\eta^3}{768\pi^3}
|f_1+f_2|^2\;.
\label{dd}
\end{eqnarray}
Substituting the results of $f_i$ into eq.~(\ref{dd}), we derive
the behavior of $R(\eta)$ as in fig.~2b, which shows a slow decrease
with $\eta$.

In order to have the full spectrum in $\eta$,
we approximate $\d\Gamma/\d\eta$ in
the range of $\eta<0.3$
by the soft pion limits of $f_i$ \cite{BLN}:
\begin{equation}
\lim_{\eta\to 0}R(\eta)=
\frac{G_F^2m^5\eta}{192\pi^3}\frac{f_{B^*}^2}{f_\pi^2}
g_{BB^*\pi}^2\;,
\label{spl}
\end{equation}
which shows a linear relation with $\eta$.
Here $f_{B^*}\approx 1.1 f_B$ \cite{N2} is
the decay constant of the $B^*$ meson, and
$g_{BB^*\pi}\approx 0.8$ \cite{YL} is the $BB^*\pi$ coupling constant.
We extrapolate eq.~(\ref{spl}) to $\eta=0.3$ as shown in fig.~2b,
and a good match between the soft pion and PQCD predictions is observed.
Certainly, this extrapolation of the soft
pion limit may not be reliable, but the
match of the two different approaches does justify our calculation
to some extent.

It is then possible to estimate the total decay rate $\Gamma$
by combining eq.~(\ref{spl}) for $\eta < 0.3$
with the PQCD predictions for $\eta > 0.3$. We obtain
$\Gamma\approx 2.4\times 10^{-11} |V_{ub}|^2$ GeV,
which corresponds to a branching ratio $0.47\times 10^2|V_{ub}|^2$
for the total width $(0.51\pm 0.02)\times 10^{-9}$ MeV of
the $B^0$ meson \cite{RPP}.
Current experimental limit on the branching ratio of $B^0\to
\pi^- l^+\nu$ is $3.3\times 10^{-4}$ \cite{O}. We then extract
the matrix element $|V_{ub}|\approx 2.7\times 10^{-3}$,
close to the value 0.003
given in the literature \cite{RPP}.


\vskip 0.5cm
We thank G.L. Lin, M. Neubert, G. Sterman and Y.P. Yao for helpful
discussions. This work was supported by the National
Science Council of R.O.C. under Grant Nos. NSC84-2112-M194-006 and
NSC84-211-M001-034.

\newpage
\cl{\large \bf Figure Captions}
\vskip 0.5cm

\noindent
{\bf Fig. 1.} Graphic representation of eq.~(\ref{qp}).
\vskip 0.5cm

\noindent
{\bf Fig. 2.} (a) Dependence of $f_1+f_2$
on the cutoff $b_c$ for (1) $\eta=0.3$,
(2) $\eta=0.4$, and (3) $\eta=1.0$.
(b) Dependence of $R(\eta)$
on $\eta$ derived from the modified PQCD formalism
(solid line) and from the soft pion theorems (dashed line).


\newpage
\begin{thebibliography}{99}
\bibitem{LR} H. Leutwyler and M. Roos, Z. phys. C25, 91 (1984).
\bibitem{N1} M. Neubert, Phys. Lett. B264, 455 (1991).
\bibitem{IW} N. Isgur and M.B. Wise, Phys. Rev. D42, 2388 (1990).
\bibitem{BLN} G. Burdman, Z. Ligeti, M. Neubert and Y. Nir, Phys. Rev
D49, 2331 (1994).
\bibitem{YS} R. Akhoury, G. Sterman and Y.-P Yao, Phys. Rev. D50, 358
(1994).
\bibitem{BS} J. Botts and G. Sterman, Nucl. Phys. B325, 62 (1989).
\bibitem{LS} H.-n. Li and G. Sterman, Nucl. Phys. B381, 129 (1992);
H.-n. Li, Phys. Rev. D48, 4243 (1993).
\bibitem{CS} J.C. Collin and D.E. Soper, Nucl. Phys. B193, 381 (1981).
\bibitem{JK} R. Jakob and P. Kroll, Phys. Lett. B315, 463 (1993);
J. Bolz, R. Jakob, P. Kroll, M. Bergmann and N.G. Stefanis,
Wuppertal Preprint WU-B-94-09.
\bibitem{C} J.C. Collins, in {\it Perturbative Quantum Chromodynamics},
ed. A.H. Mueller (World Scientific, Singapore, 1989).
\bibitem{CZ} V.L. Chernyak and A.R. Zhitnitsky,
Nucl. Phys. B201, 492 (1982); Phys. Rep. 112, 173 (1984).
\bibitem{S} F. Schlumpf, SLAC-PUB-6335.
\bibitem{BLS} C. Bernard, J. Labrenz and A. Soni, Nucl. Phys.
B(Proc.)30, 465 (1993).
\bibitem{N2} M. Neubert, Phys. Rev. D46, 1076 (1992).
\bibitem{YL} T.M. Yan {\it et al.}, Phys. Rev. D46, 1148 (1992).
\bibitem{RPP} Review of Particle Properties, Phys. Rev. D45 (1992).
\bibitem{O} B. Ong {\it et al.} (CLEO collaboration), Phys. Rev. Lett.
70, 18 (1993).
\end{thebibliography}
\end{document}